# The nonlinear field theory I

## The Dirac electron theory as approximation of the nonlinear electrodynamics

Alexander G. Kyriakos

Saint-Petersburg State Institute of Technology,
St.Petersburg, Russia.

Present address: Athens, Greece, e-mail: agkyriak@yahoo.com

**Abstract**
 In the present paper it is shown that the Dirac electron theory is the approximation of the special nonlinear electromagnetic field theory.

## 1.0. Introduction.

### 1.1. The Dirac equation form of the Maxwell equations

It is well known that Maxwell's and Dirac's equations can be written in very similar forms [1,2,3]. For instance the Maxwell equations without the current can be represented [3] in the form**:**

$$\frac{i}{c}\frac{\partial \vec{F}}{\partial t} = (\hat{\vec{S}} \cdot \hat{\vec{p}})\vec{F}, \qquad (1.1)$$

which is called the Dirac's form of the Maxwell equations because they are similar to the Dirac electron equation [2,20]:

$$i\frac{\partial \psi}{\partial t} = \hat{H}\psi, \qquad (1.2)$$

(here $\hat{\vec{p}}$ is the momentum operator, $\vec{F} = \vec{E} + i\vec{H}$, $\vec{F}^* = \vec{E} - i\vec{H}^*$ are the so-called Cramers 6-vectors, $\vec{E}$ and $\vec{H}$ are the 3-vectors of the electromagnetic field and $\hat{\vec{S}}$ are the photon 3x3 spin matrices, $\hat{H} = \hat{\vec{\alpha}} \cdot \hat{\vec{p}} + \hat{\beta} m$ is the Hamiltonian of the Dirac electron equation and $\psi$ is the Dirac wave function). Actually it is not difficult to see that in the case $m = 0$ the equations (1.1) and (1.2) are equivalent. (The function $\vec{F}$ can also be written in the spinor or bispinor matrix form [2]).

But [4] "the conventional view is that spin 1 and spin ½ particles belong to distinct irreducible representations of the Poincare group, so that there should be no connection between the Maxwell and Dirac equations describing of these particles". Actually there are two serious differences between the Dirac and the Maxwell theories [4]: at first the Dirac theory is exclusively a complex theory while the Maxwell theory is a real theory; secondly the Dirac bispinors contain four functions, while the electromagnetic field in the general case contains six functions. Obviously, in this case [4] "there is no physically meaningful way to transform

Maxwell's and Dirac's equations into each other". That is why the full representation of the Dirac theory in the Maxwell form is absent until today.

The question arises is there a possibility of the fully electromagnetic representation of the Dirac theory? This question can also be formulated otherwise as follows: can the vectorial Maxwell theory produce the spinor Dirac theory?

In the present paper we attempt answer these questions. We will show that such a possibility exists only in the case, when Dirac's bispinor is identified with the fields of the nonlinear electromagnetic wave.

## 1.2. Brief review of the present nonlinear field theories

### 1.2.1. Nonlinear electron theories

The most well known the non-linear field theories are the M. Born - L. Infeld [5] and the W. Heisenberg – H. Euler electrodynamics [6].

M. Born and L. Infeld proceeded from the idea of a limited value of the electromagnetic field strength of the electron. This reason and some others led them to the following Lagrangian of the nonlinear electrodynamics in the vacuum:

$$L_{BI} = \frac{1}{4\pi a^2}\left(1-\sqrt{1+a^2(\vec{E}^2-\vec{B}^2)-a^4(\vec{E}\cdot\vec{B})^2}\right), \qquad (1.3)$$

where $a = const$.

In the case of weak fields, the Lagrangian of the Born-Infeld nonlinear electrodynamics can be expanded into the small parameters $a^2 E^2 \ll 1$ and $a^2 B^2 \ll 1$:

$$L_{BI} \approx -\frac{1}{8\pi}(\vec{E}^2-\vec{B}^2)+\frac{a^2}{32\pi}\left[(\vec{E}^2-\vec{B}^2)^2+4(\vec{E}\cdot\vec{B})^2\right], \qquad (1.4)$$

The nonlinear electrodynamics of W. Heisenberg and H. Euler appeared on the base of the quantum electrodynamics effect of the electron-positron vacuum polarization. If the electromagnetic fields are not strong, the vacuum electromagnetic nonlinear field Lagrangian should have the following form [6]:

$$L_{BI} = -\frac{1}{8\pi}(\vec{B}^2-\vec{H}^2)+\frac{\alpha_q^2}{360\pi^2}\left[(\vec{B}^2-\vec{H}^2)^2+7(\vec{E}\cdot\vec{H})^2\right], \qquad (1.5)$$

where $\alpha_q = e^2/\hbar c \approx 1/137$ is the fine structure constant.

Now we will briefly consider the important for our theme results of the Born - Infeld theory [5,7] for the most important case of an electrostatic field of a spherical electron. For an electrical induction we obtain here:

$$D_r = \frac{e\vec{r}}{r^3}, \qquad (1.6)$$

i.e., *from point of view of the D-field, the electron should be considered as point*.

For the electric field (E-field) we obtain:

$$\vec{E}_r = \frac{\vec{D}_r}{\sqrt{1+\frac{D_r^2}{E_0^2}}} = \frac{e\vec{r}}{r\sqrt{r^4+r_0^4}}, \qquad (1.7)$$

where $r_0 = \sqrt{\frac{e}{E_0}}$, i.e. *from point of view of the electric field (E-field) the electron is not a point*. This is very important specificity of the nonlinear theory with comparison to the linear theory.

As it is known, the two types of fields and the two definitions of the charge density, corresponding to them, are also described by the theory of the dielectrics. The value:

$$\varepsilon = \frac{D}{E} = \sqrt{1 + \frac{r_0^4}{r^4}} \ , \tag{1.8}$$

which is here a function of the position, can be considered as a "dielectric permeability" of the vacuum. The vacuum polarization effects arise, when in some area of space there is a very strong field. It is important to highlight, that according to this fact the various effects of the dispersive type should be observed: effect of birefringence, bending of a wave, etc. An experimental method to detect the vacuum polarization effects was considered in [8,9]. The theoretical description of the nonlinear effects on light propagation was studied long before in [10,11]. Other new results on vacuum polarization phenomena in nontrivial vacuum, including curved space-time, can be found in the works [12-17]. The first experimental verification of light-by-light scattering in the vacuum was obtained by Burke, Feld, *et al.*, [18].

*1.2.2. Non-linear generalisation of the Dirac electron theory*

The interconversion of the elementary particles can be considered as a problem of the nonlinear quantum electrodynamics [2]. The simplest kind of the nonlinear generalization of the Dirac equation looks like [19]:

$$\left(D_0 + \lambda \psi^2\right) \psi = 0, \tag{1.9}$$

where $D_o = \left(\hat{\varepsilon} - c\hat{\alpha} \ \hat{\vec{p}} - \hat{\beta} \ mc^2\right)$ is the operator of the Dirac equation.

It is necessary to note, that the above nonlinear quantum theories have the same disadvantage with the nonlinear classical theories: the choice of the Lagrangian has not got any support in the theory, except for some general requirements of the symmetry.

Below we will show that the Lagrangian of the nonlinear electron theory appears automatically, when the Dirac (bi)spinors are identified with the nonlinear electromagnetic waves. We will also show that the Maxwell and Dirac theories are both the approximation of the same nonlinear field theory.

## 2.0. The electrodynamics form of the Dirac equation without mass

Let's recall the usual quantum form of the Dirac electron equation. There are two bispinor Dirac equation forms [2,20]:

$$\left[\left(\hat{\alpha}_o \hat{\varepsilon} + c\hat{\vec{\alpha}} \cdot \hat{\vec{p}}\right) + \hat{\beta} \ mc^2\right] \psi = 0 , \tag{2.1}$$

$$\psi^+ \left[\left(\hat{\alpha}_o \hat{\varepsilon} - c\hat{\vec{\alpha}} \cdot \hat{\vec{p}}\right) - \hat{\beta} \ mc^2\right] = 0 , \tag{2.2}$$

which correspond to the two signs of the relativistic expression of the energy of the electron:

$$\varepsilon = \pm \sqrt{c^2 \vec{p}^2 + m^2 c^4} \ , \tag{2.3}$$

where $\hat{\varepsilon} = i\hbar \dfrac{\partial}{\partial t}, \hat{\vec{p}} = -i\hbar \vec{\nabla}$ are the operators of the energy and momentum, $\varepsilon, \vec{p}$ are the electron energy and momentum, $c$ is the light velocity, $m$ is the electron mass, and $\hat{\alpha}_o = \hat{1}$, $\hat{\vec{\alpha}}$, $\hat{\alpha}_4 \equiv \hat{\beta}$ are the Dirac matrices:

$$\hat{\alpha}_1 = \begin{pmatrix} 0 & 0 & 0 & 1 \\ 0 & 0 & 1 & 0 \\ 0 & 1 & 0 & 0 \\ 1 & 0 & 0 & 0 \end{pmatrix}, \quad \hat{\alpha}_2 = \begin{pmatrix} 0 & 0 & 0 & -i \\ 0 & 0 & i & 0 \\ 0 & -i & 0 & 0 \\ i & 0 & 0 & 0 \end{pmatrix}, \quad \hat{\alpha}_3 = \begin{pmatrix} 0 & 0 & 1 & 0 \\ 0 & 0 & 0 & -1 \\ 1 & 0 & 0 & 0 \\ 0 & -1 & 0 & 0 \end{pmatrix}, \quad \vec{\alpha}_4 = \begin{pmatrix} 1 & 0 & 0 & 0 \\ 0 & 1 & 0 & 0 \\ 0 & 0 & -1 & 0 \\ 0 & 0 & 0 & -1 \end{pmatrix},$$

$\psi$ is the wave function $\psi = \begin{pmatrix} \psi_1 \\ \psi_2 \\ \psi_3 \\ \psi_4 \end{pmatrix}$ called bispinor, $\psi^+$ is the Hermitian-conjugate wave function.

Now we will consider the derivation of the electrodynamics form of the Dirac equations without mass term. Let us consider the plane electromagnetic wave moving on $y$ - axis. In the general case it has two polarizations and contains the following field vectors:

$$E_x, E_z, H_x, H_z, \tag{2.4}$$

(As it is known, for all transformations the relation $E_y = H_y = 0$ takes place, so that there are always only four components (2.4) as in the Dirac $\psi$ -function).

Let's enter the electromagnetic wave fields as the Dirac bispinor matrix:

$$\psi = \begin{pmatrix} E_x \\ E_z \\ iH_x \\ iH_z \end{pmatrix}, \quad \psi^+ = (E_x \quad E_z \quad -iH_x \quad -iH_z), \tag{2.5}$$

(For all other directions of the electromagnetic waves the matrices choice will be considered in the following paper).

Then the Klein-Gordon equation [20] without mass can be written as:

$$\left(\hat{\varepsilon}^2 - c^2 \hat{\vec{p}}^2\right)\psi = 0, \tag{2.6}$$

Using (2.5), we can prove that (2.6) is also the equation of the electromagnetic wave, moving along the $y$ - axis. The equation (2.6) can also be written in the following form:

$$\left[\left(\hat{\alpha}_o \hat{\varepsilon}\right)^2 - c^2 \left(\hat{\vec{\alpha}} \cdot \hat{\vec{p}}\right)^2\right] \psi = 0, \tag{2.7}$$

In fact, taking into account that

$$\left(\hat{\alpha}_o \hat{\varepsilon}\right)^2 = \hat{\varepsilon}^2, \quad \left(\hat{\vec{\alpha}} \cdot \hat{\vec{p}}\right)^2 = \hat{\vec{p}}^2, \tag{2.8}$$

we see that the equations (2.6) and (2.7) are equivalent.

Factorizing (2.7) and multiplying it from the left by the Hermitian-conjugate function $\psi^+$ we get:

$$\psi^+ \left(\hat{\alpha}_o \hat{\varepsilon} - c\hat{\vec{\alpha}} \cdot \hat{\vec{p}}\right) \left(\hat{\alpha}_o \hat{\varepsilon} + c\hat{\vec{\alpha}} \cdot \hat{\vec{p}}\right) \psi = 0, \tag{2.9}$$

The equation (2.9) may be disintegrated in two Dirac equations without mass:

$$\psi^+ \left(\hat{\alpha}_o \hat{\varepsilon} - c\hat{\vec{\alpha}} \cdot \hat{\vec{p}}\right) = 0, \tag{2.10}$$

$$\left(\hat{\alpha}_o \hat{\varepsilon} + c\hat{\vec{\alpha}} \cdot \hat{\vec{p}}\right) \psi = 0, \tag{2.11}$$

It is not difficult to show (using (2.5) that the equations (2.10) and (2.11) are also the Maxwell equations of the plane electromagnetic waves.

Now we will consider the question about the appearance of the mass in the Dirac equation.

## 3.0. The appearance of the mass term

The disintegration (2.6) in (2.10) and (2.11) can be compared with the typical transformation of the massless quantum of an electromagnetic wave $\gamma$ in two massive particles (electron-positron) $e^-, e^+$:

$$\gamma \to e^+ + e^-, \tag{3.1}$$

Then the question arises: which mathematical transformation can turn the equations (2.10) and (2.11) into the equations (2.1) and (2.2)?

We will show that in quantum form this transformation corresponds to the transition of the Dirac wave function (i.e. bispinor) from the flat Euclidean to the Riemann curvilinear space.

### 3.1. The generalization of the Dirac equation on the Riemann geometry

The generalization of the Dirac equation on the Riemann geometry is connected with the parallel transport and covariant differentiation of the spinor in the curvilinear space. These problems were considered for the first time in the articles [21] and further in the articles [22]. We will use the most important results of this theory below.

For the generalization of the Dirac equation on the Riemann geometry it is necessary [21,22] to replace the usual derivative $\partial_\mu \equiv \partial/\partial x_\mu$ (where $x_\mu$ is the co-ordinates in the 4-space) with the covariant derivative:

$$D_\mu = \partial_\mu + \Gamma_\mu, \tag{3.2}$$

where $\mu = 0, 1, 2, 3$ are the summing indices and $\Gamma_\mu$ is the analogue of Christoffel's symbols in the case of the spinor theory (called Ricci connection coefficients or the coefficient of the parallel transport of the spinor). In the theory it is shown [21] that $\hat{\alpha}_\mu \Gamma_\mu = \hat{\alpha}_i p_i + i\hat{\alpha}_0 p_0$, where $p_i$ and $p_0$ are not the operators, but the physical values.

Thus, using (3.2) we obtain from (2.10) and (2.11):

$$\alpha^\mu D_\mu \psi = \alpha^\mu (\partial_\mu + \Gamma_\mu)\ \psi = 0,$$

When a spinor moves along the beeline, all of the $\Gamma_\mu = 0$ and we have a usual derivative. But if a spinor moves along the curvilinear trajectory, then not all of the $\Gamma_\mu$ are equal to zero and a supplementary term appears. Typically, the last one is not the derivative, but it is equal to the product of the spinor itself with some coefficient $\Gamma_\mu$. It is not difficult to show that the supplementary term contains a mass.

Since, according to general theory [21,22], the increment in spinor $\Gamma_\mu$ has the form and the dimension of the 4-vector of the energy-momentum, it is logical to identify $\Gamma_\mu$ with 4-vector of energy-momentum of the photon electromagnetic field:

$$\Gamma_\mu = \{\varepsilon_p, c\vec{p}_p\}, \tag{3.3}$$

where $\varepsilon_p$ and $p_p$ are the electromagnetic wave energy and momentum (below we will prove this supposition from other point of view). Then equations (2.10) and (2.11) in the curvilinear space will have the form:

$$\psi^+ \left[ \left( \hat{\alpha}_o \hat{\varepsilon} - c\hat{\vec{\alpha}} \cdot \hat{\vec{p}} \right) - \left( \hat{\alpha}_o \varepsilon_p - c\hat{\vec{\alpha}} \cdot \vec{p}_p \right) \right] = 0, \tag{3.4}$$

$$\left[ \left( \hat{\alpha}_o \hat{\varepsilon} + c\hat{\vec{\alpha}} \cdot \hat{\vec{p}} \right) + \left( \hat{\alpha}_o \varepsilon_p + c\hat{\vec{\alpha}} \cdot \vec{p}_p \right) \right] \psi = 0, \tag{3.5}$$

In the case of the reaction (3.1) according to the energy conservation law we can formally write:

$$\hat{\alpha}_o \varepsilon_p \pm c\hat{\vec{\alpha}} \cdot \vec{p}_p = \mp \hat{\beta} m c^2, \tag{3.6}$$

Substituting (3.6) in (3.4) and (3.5) we will arrive at the usual kind of the Dirac equation with the mass:

$$\psi^+ \left[ \left( \hat{\alpha}_o \hat{\varepsilon} - c\hat{\vec{\alpha}} \cdot \hat{\vec{p}} \right) - \hat{\beta} m c^2 \right] = 0, \tag{3.7}$$

$$\left[ \left( \hat{\alpha}_o \hat{\varepsilon} + c\hat{\vec{\alpha}} \cdot \hat{\vec{p}} \right) + \hat{\beta} m c^2 \right] \psi = 0, \tag{3.8}$$

We will find now, which value corresponds to the mass term in the electromagnetic form of the Dirac equation.

### 3.2. Electrodynamics form of the Dirac equation with mass

Consider two Hermitian-conjugate equations, corresponding to the minus sign of the expression (2.3):

$$\left[ \left( \hat{\alpha}_o \hat{\varepsilon} + c\hat{\vec{\alpha}} \cdot \hat{\vec{p}} \right) + \hat{\beta} mc^2 \right] \psi = 0, \tag{3.9}$$

$$\psi^+ \left[ \left( \hat{\alpha}_o \hat{\varepsilon} + c\hat{\vec{\alpha}} \cdot \hat{\vec{p}} \right) + \hat{\beta} mc^2 \right] = 0, \tag{3.10}$$

Using (2.5), from (3.9) and (3.10) we obtain:

$$\begin{cases} \dfrac{1}{c}\dfrac{\partial E_x}{\partial t} - \dfrac{\partial H_z}{\partial y} + i\dfrac{\omega}{c} E_x = 0, \\ \dfrac{1}{c}\dfrac{\partial E_z}{\partial t} + \dfrac{\partial H_x}{\partial y} + i\dfrac{\omega}{c} E_z = 0, \\ \dfrac{1}{c}\dfrac{\partial H_x}{\partial t} + \dfrac{\partial E_z}{\partial y} - i\dfrac{\omega}{c} H_x = 0, \\ \dfrac{1}{c}\dfrac{\partial H_z}{\partial t} - \dfrac{\partial E_x}{\partial y} - i\dfrac{\omega}{c} H_z = 0, \end{cases} \tag{3.11} \qquad \begin{cases} \dfrac{1}{c}\dfrac{\partial E_x}{\partial t} - \dfrac{\partial H_z}{\partial y} - i\dfrac{\omega}{c} E_x = 0, \\ \dfrac{1}{c}\dfrac{\partial E_z}{\partial t} + \dfrac{\partial H_x}{\partial y} - i\dfrac{\omega}{c} E_z = 0, \\ \dfrac{1}{c}\dfrac{\partial H_x}{\partial t} + \dfrac{\partial E_z}{\partial y} + i\dfrac{\omega}{c} H_x = 0, \\ \dfrac{1}{c}\dfrac{\partial H_z}{\partial t} - \dfrac{\partial E_x}{\partial y} + i\dfrac{\omega}{c} H_z = 0, \end{cases} \tag{3.12}$$

where $\omega = \dfrac{mc^2}{\hbar}$. The equations (3.11) and (3.12) are Maxwell equations with imaginary currents, which differ by the directions. These currents have the following form:

$$\vec{j}_e = i\dfrac{\omega}{4\pi}\vec{E}, \quad \vec{j}_m = i\dfrac{\omega}{4\pi}\vec{H}, \tag{3.13}$$

and it is interesting that together with the electric current $j_e$ the magnetic current $j_m$ also exists here. The last one must be equal zero according to the Maxwell theory [23], but its existence according to Dirac doesn't contradict to the quantum theory (see the Dirac theory of the magnetic monopole [24]).

Thus, the term, which in the Dirac equation contains the electron mass, corresponds to the term that in the Maxwell equation contains the imaginary electric and the complex "magnetic" current.

Now we will consider the origin of the appearance of the current-mass term in the electromagnetic form.

### 3.3. The displacement ring current

We will consider the Maxwell equations without current (2.10) or (2.11), as the equations of the initial photon field. Then, the reaction (3.1) can be formally understood in such a way, that while moving through the nucleus field, the electromagnetic wave fields may undergo a transformation and convert into the electron - positron pair.

We will show that the reason why the current-mass term appears in the equations (2.10) and (2.11) is the transition of the initial electromagnetic wave field from the linear to the curvilinear trajectory and that this term is the supplementary Maxwell displacement current.

Let the plane-polarized wave, which have the field vectors $(E_x, H_z)$, be twirled with some radius $r_p$ in the plane $(X', O', Y')$ of a fixed co-ordinate system $(X', Y', Z', O')$ so that $E_x$ is parallel to the plane $(X', O', Y')$ and $H_x$ is perpendicular to it.

According to Maxwell [23] the displacement current is defined by the equation:

$$\vec{j}_{dis} = \frac{1}{4\pi} \frac{\partial \vec{E}}{\partial t}, \tag{3.14}$$

The above electrical field vector $\vec{E}$, which moves along the curvilinear trajectory (let it have the direction from the centre), can be written in the form:

$$\vec{E} = -E\,\vec{n}, \tag{3.15}$$

where $E = |\vec{E}|$ and $\vec{n}$ is the normal unit-vector of the curve (having direction to the center). The derivative of $\vec{E}$ with respect to $t$ can be represented as:

$$\frac{\partial \vec{E}}{\partial t} = -\frac{\partial E}{\partial t}\vec{n} - E\frac{\partial \vec{n}}{\partial t}, \tag{3.16}$$

Here the first term has the same direction as $\vec{E}$. The existence of the second term shows that at the wave twirling the supplementary displacement current appears. It is not difficult to show that it has a direction, tangential to the ring:

$$\frac{\partial \vec{n}}{\partial t} = -\frac{\upsilon_p}{r_p}\vec{\tau}, \tag{3.17}$$

where $\vec{\tau}$ is the tangential unit-vector, $\upsilon_p \equiv c$ is the electromagnetic wave velocity. Thus, the displacement current of the ring wave can be written in the form:

$$\vec{j}_{dis} = -\frac{1}{4\pi} \frac{\partial E}{\partial t}\vec{n} + \frac{1}{4\pi}\omega_p E\,\vec{\tau}, \tag{3.18}$$

where $\omega_p = \frac{\upsilon_p}{r_p}$ is the angular velocity, $\vec{j}_n = \frac{1}{4\pi}\frac{\partial E}{\partial t}\vec{n}$ and $\vec{j}_\tau = \frac{\omega_p}{4\pi}E\,\vec{\tau}$ are the normal and tangent components of the current of the twirled electromagnetic wave, correspondingly. Thus:

$$\vec{j}_{dis} = \vec{j}_n + \vec{j}_\tau, \tag{3.19}$$

The currents $\vec{j}_n$ and $\vec{j}_\tau$ are always mutually perpendicular, so that we can write them in the complex form: $j_{dis} = j_n + i j_\tau$, where $j_\tau = \frac{\omega_p}{4\pi}E$.

Thus, as we see, the transition of the initial electromagnetic wave from the linear to the curvilinear trajectory corresponds to the production of the Dirac bispinor theory. We can prove it through the analysis of the free electron equation solution.

## 4.0. Electromagnetic form of the free electron equation solution

In accordance with the above results the electromagnetic form of the solution of the Dirac free electron equation must be a twirled electromagnetic wave. Let us show that this supposition is actually correct.

From the above point of view for the $y$-direction photon two solutions must exist:
1) for the wave, twirled around the $OZ$-axis

$$^{oz}\psi = \begin{pmatrix} E_x \\ 0 \\ 0 \\ iH_z \end{pmatrix} = \begin{pmatrix} \psi_1 \\ 0 \\ 0 \\ \psi_4 \end{pmatrix}, \quad (4.1)$$

and 2) for the wave, twirled around the $OX$-axis

$$^{ox}\psi = \begin{pmatrix} 0 \\ E_z \\ iH_x \\ 0 \end{pmatrix} = \begin{pmatrix} 0 \\ \psi_2 \\ \psi_3 \\ 0 \end{pmatrix}, \quad (4.2)$$

Let us compare (4.1) and (4.2) with the Dirac theory solutions.
It is known [2,20] that the solution of the Dirac free electron equation (2.1) has the form of the plane wave:

$$\psi_j = B_j \exp\left(-\frac{i}{\hbar}(\varepsilon t - \vec{p}\vec{r})\right) \quad (4.3)$$

where $j = 1, 2, 3, 4$; $B_j = b_j e^{i\phi}$; the amplitudes $b_j$ are the numbers and $\phi$ is the initial wave phase. The functions (4.3) are the eigenfunctions of the energy-momentum operators, where $\varepsilon$ and $\vec{p}$ are the energy-momentum eigenvalues. Here for each $\vec{p}$, the energy $\varepsilon$ has either positive or negative values of equation (2.3) correspondingly.

For $\varepsilon_+$ we have two linear-independent set of four orthogonal normalising amplitudes:

$$1)\ B_1 = -\frac{cp_z}{\varepsilon_+ + mc^2},\ B_2 = -\frac{c(p_x + ip_y)}{\varepsilon_+ + mc^2},\ B_3 = 1,\ B_4 = 0, \quad (4.4)$$

$$2)\ B_1 = -\frac{c(p_x - ip_y)}{\varepsilon_+ + mc^2},\ B_2 = \frac{cp_z}{\varepsilon_+ + mc^2},\ B_3 = 0,\ B_4 = 1, \quad (4.5)$$

and for $\varepsilon_-$:

$$3)\ B_1 = 1,\ B_2 = 0,\ B_3 = \frac{cp_z}{-\varepsilon_- + mc^2},\ B_4 = \frac{c(p_x + ip_y)}{-\varepsilon_- + mc^2}, \quad (4.6)$$

$$4)\ B_1 = 0,\ B_2 = 1,\ B_3 = \frac{c(p_x - ip_y)}{-\varepsilon_- + mc^2},\ B_4 = -\frac{cp_z}{-\varepsilon_- + mc^2}, \quad (4.7)$$

Let's analyze these solutions.
**1)** The existing of two linear independent solutions corresponds with two independent orientations of the electromagnetic wave vectors and gives the unique logic explanation for this fact.
**2)** Since $\psi = \psi(y)$, we have $p_x = p_z = 0$, $p_y = mc$ and for the field vectors we obtain: from (4.4) and (4.5) for "positive" energy

$$B_+^{(1)} = \begin{pmatrix} 0 \\ b_2 \\ b_3 \\ 0 \end{pmatrix} \cdot e^{i\phi},\ B_+^{(2)} = \begin{pmatrix} b_1 \\ 0 \\ 0 \\ b_4 \end{pmatrix} \cdot e^{i\phi}, \quad (4.8)$$

and from (4.6) and (4.7) for "negative" energy:

$$B_-^{(1)} = \begin{pmatrix} b_1 \\ 0 \\ 0 \\ b_4 \end{pmatrix} \cdot e^{i\phi}, \quad B_-^{(2)} = \begin{pmatrix} 0 \\ b_2 \\ b_3 \\ 0 \end{pmatrix} \cdot e^{i\phi}, \qquad (4.9)$$

which exactly correspond to (4.1) and (4.2).

**3)** Calculate the correlations between the components of the field vectors. Putting $\phi = \dfrac{\pi}{2}$ for $\varepsilon_+ = mc^2$ and $\varepsilon_- = -mc^2$ we obtain correspondingly:

$$B_+^{(1)} = \begin{pmatrix} 0 \\ \frac{1}{2} \\ i \cdot 1 \\ 0 \end{pmatrix}, \quad B_+^{(2)} = \begin{pmatrix} -\frac{1}{2} \\ 0 \\ 0 \\ i \cdot 1 \end{pmatrix}, \qquad (4.10)$$

$$B_-^{(1)} = \begin{pmatrix} i \cdot 1 \\ 0 \\ 0 \\ -\frac{1}{2} \end{pmatrix}, \quad B_-^{(2)} = \begin{pmatrix} 0 \\ i \cdot 1 \\ \frac{1}{2} \\ 0 \end{pmatrix}, \qquad (4.11)$$

The imaginary unit in these solutions indicates that the field vectors $\vec{E}$ and $\vec{H}$ are mutualy orthogonal. Also we see that the electric field amplitude is two times less, than the magnetic field amplitude. This fact corresponds to the electromagnetic spinor contrary to the linear wave of the Maxwell theory, where the field vectors $\vec{E}$ and $\vec{H}$ are equal. It can be show that this relation is needed for the electron stability. (Note also, that the appearance of minus sign before the electric field vector in $B_+^{(1)}$ and $B_-^{(1)}$ corrects automatically our mistake in (4.1)).

**4)** It is easy to show that the electromagnetic form of the solution of the Dirac equation is the standing wave. Really in case of the twirled wave we have $\vec{p} \cdot \vec{r} = 0$ and instead (4.3) we obtain:

$$\psi_j = b_j \exp\left(-\frac{i}{\hbar}\varepsilon\, t\right), \qquad (4.12)$$

**5)** According with the Euler formula $e^{i\varphi} = \cos\varphi + i\sin\varphi$ the solution of the Dirac equation (4.12) describes a circle.

It is appropriate to note here that in case when the function (2.5) is the solution of the Dirac equation in the electromagnetic form we can name it "electromagnetic bispinor". In other words the electromagnetic bispinor is the electromagnetic wave, moving along the curvilinear trajectory.

From this follows that the Dirac electron equation in the electromagnetic form must be the nonlinear electromagnetic field equation. Below, we will clarify the explicit form of this equation.

## 5.0. Nonlinear electrodynamics representation of the Dirac electron theory

### 5.1. The nonlinear Dirac equation

The stability of twirled photon is possible only by the photons parts self-action. We could introduce the self-action of the fields to the Dirac equation like the external interaction is introduced to the quantum field theory equations, putting herewith the photon mass equal to zero [2,20]. But this equation may be obtained more easily, using the relation (3.6), for example, the expression

$$\hat{\beta}\, mc^2 = -\left(\varepsilon_p - c\hat{\vec{\alpha}} \cdot \vec{p}_p\right), \tag{5.1}$$

By substituting (5.1) to the Dirac electron equation we obtain the nonlinear integral equation:

$$\left[\hat{\alpha}_0(\hat{\varepsilon} - \varepsilon_p) + c\hat{\vec{\alpha}} \cdot (\hat{\vec{p}} - \vec{p}_p)\right]\psi = 0, \tag{5.2}$$

which is, as we suppose, the common form of the nonlinear equation, which describes the electron in both quantum and concurrent electromagnetic forms. Let us show that in the approximate form it gives the forms that are known in the modern theory.

In the electromagnetic form we have:

$$\varepsilon_p = \int_0^\tau U\, d\tau = \frac{1}{8\pi}\int_0^\tau \left(\vec{E}^2 + \vec{H}^2\right) d\tau, \tag{5.3}$$

$$\vec{p}_p = \int_0^\tau \vec{g}\, d\tau = \frac{1}{c^2}\int_0^\tau \vec{S}\, d\tau = \frac{1}{4\pi}\int_0^\tau \left[\vec{E}\times\vec{H}\right] d\tau, \tag{5.4}$$

where in the general case the upper limit $\tau$ is equal to infinity.

Let's show the connection of the equation (5.2) with known nonlinear spinor equation. Consider the approximate form of this equation. Since the main part of the electron energy consists of some finite volume, we can writ

$$\tau \approx \Delta\tau = \varsigma\, r_p^3, \tag{5.5}$$

where $\varsigma$ is a constant. Using the quantum form of $U$ and $\vec{S}$:

$$U = \frac{1}{8\pi}\, \psi^+ \hat{\alpha}_0 \psi, \tag{5.6}$$

$$\vec{S} = -\frac{c}{8\pi}\, \psi^+ \hat{\vec{\alpha}}\, \psi = c^2 \vec{g}, \tag{5.7}$$

and taking in to account that the free electron Dirac equation solution is the plane wave:

$$\psi = \psi_0 \exp[i(\omega t - ky)], \tag{5.8}$$

we can write (5.6) and (5.7) in the next approximate form:

$$\varepsilon_p = U\, \Delta\tau = \frac{\Delta\tau}{8\pi}\psi^+ \hat{\alpha}_0 \psi, \tag{5.9}$$

$$\vec{p}_p = \vec{g}\, \Delta\tau = \frac{1}{c^2}\vec{S}\, \Delta\tau = -\frac{\Delta\tau}{8\pi\, c}\, \psi^+ \hat{\vec{\alpha}}\, \psi, \tag{5.10}$$

By substituting (5.9) and (5.10) into (5.2) and taking into account (5.5), we obtain the following approximate equation:

$$\frac{\partial \psi}{\partial t} - c\hat{\vec{\alpha}}\vec{\nabla}\psi + i\frac{\varsigma}{8\pi\, \hbar c}\cdot r_p^3 \left(\psi^+\hat{\alpha}_0\psi - \hat{\vec{\alpha}}\psi^+\hat{\vec{\alpha}}\psi\right)\psi = 0, \tag{5.11}$$

It is not difficult to see that the equation (5.11) is the nonlinear equation of the same type, which was investigated by Heisenberg et. al. [19] and which played for a while the role of the unitary field theory equation. Contrary to the last one, the equation (5.11) is obtained in a logical and correct way and the self-action constant $r_p$ appeared in (5.11) automatically.

### 5.2. The Lagrangian density of the nonlinear Dirac equation

The Lagrangian density of the linear Dirac equation in quantum form is [20]:

$$L_D = \psi^+\left(\hat{\varepsilon} + c\hat{\vec{\alpha}}\, \hat{\vec{p}} + \hat{\beta}\, mc^2\right)\psi, \tag{5.12}$$

or in the electromagnetic form:

$$L_D = \frac{\partial U}{\partial t} + \operatorname{div} \vec{S} - i\frac{\omega}{8\pi}(\vec{E}^2 - \vec{H}^2), \tag{5.12'}$$

The Lagrangian density of nonlinear equation is not difficult to obtain from the Lagrangian density of the linear Dirac equation [2,20] using the method by which we found the nonlinear equation. By substituting (5.1) in (5.12) we obtain:

$$L_N = \psi^+(\hat{\varepsilon} - c\hat{\vec{\alpha}} \cdot \hat{\vec{p}})\psi + \psi^+(\varepsilon_p - c\hat{\vec{\alpha}} \cdot \vec{p}_p)\psi, \tag{5.13}$$

We suppose that the expression (5.13) represents the common form of the Lagrangian density of the nonlinear twirled electromagnetic wave equation.

Using (5.9) and (5.10) we can represent (5.12) in the approximate quantum form:

$$L_N = i\hbar\left[\frac{\partial}{\partial t}\left[\frac{1}{2}(\psi^+\psi)\right] - c\operatorname{div}(\psi^+\hat{\vec{\alpha}}\psi)\right] + \frac{\Delta\tau}{8\pi}\left[(\psi^+\psi)^2 - (\psi^+\hat{\vec{\alpha}}\psi)^2\right], \tag{5.14}$$

By the normalizing $\psi$-function by the expression $L'_N = \frac{1}{8\pi\, mc^2}L_N$ and transforming (5.13) in the electrodynamics form, using equations (5.3) and (5.4), we will obtain from (5.13) the following approximate electromagnetic form:

$$L'_N = i\frac{\hbar}{2mc^2}\left(\frac{\partial U}{\partial t} + \operatorname{div} \vec{S}\right) + \frac{\Delta\tau}{mc^2}(U^2 - c^2\vec{g}^2), \tag{5.14'}$$

It is not difficult to transform the second summand, using the known electrodynamics transformation:

$$(8\pi)^2(U^2 - c^2\vec{g}^2) = (\vec{E}^2 + \vec{H}^2)^2 - 4(\vec{E}\times\vec{H})^2 = (\vec{E}^2 - \vec{H}^2)^2 + 4(\vec{E}\cdot\vec{H})^2, \tag{5.15}$$

Thus, taking into account that $L_D = 0$ and using (5.12') and (5.15), we obtain from (5.14) the following expression:

$$L'_N = \frac{1}{8\pi}(\vec{E}^2 - \vec{H}^2) + \frac{\Delta\tau}{(8\pi)^2 mc^2}\left[(\vec{E}^2 - \vec{H}^2)^2 + 4(\vec{E}\cdot\vec{H})^2\right], \tag{5.16}$$

As we see, the approximate form of the Lagrangian density of the nonlinear equation of the twirled electromagnetic wave contains only the invariants of the Maxwell theory and is similar to the known Lagrangian density of the photon-photon interaction [2,25].

Let's now analyze the quantum form of the Lagrangian density (5.16). The equation (5.12) can be written in the form:

$$L_Q = \psi^+\hat{\alpha}_\mu\partial_\mu\psi + \frac{\Delta\tau}{8\pi}\left[(\psi^+\hat{\alpha}_0\psi)^2 - (\psi^+\hat{\vec{\alpha}}\,\psi)^2\right], \tag{5.17}$$

It is not difficult to see that the electrodynamics correlation (5.15) in quantum form has the known form of the Fierz correlation [26]:

$$(\psi^+\hat{\alpha}_0\psi)^2 - (\psi^+\hat{\vec{\alpha}}\,\psi)^2 = (\psi^+\hat{\alpha}_4\psi)^2 + (\psi^+\hat{\alpha}_5\psi)^2, \tag{5.18}$$

Using (5.18) from (5.17) we obtain:

$$L_Q = \psi^+\hat{\alpha}_\mu\partial_\mu\psi + \frac{\Delta\tau}{8\pi}\left[(\psi^+\hat{\alpha}_4\psi)^2 - (\psi^+\hat{\alpha}_5\psi)^2\right], \tag{5.19}$$

The Lagrangian density (5.19) coincides with the Nambu and Jona-Lasinio Lagrangian density [27], which is the Lagrangian density of the relativistic superconductivity theory. As it is known this Lagrangian density is used for the solution of the problem of the elementary particles mass appearance by the mechanism of the vacuum symmetry spontaneous breakdown (it corresponds also to the Cooper's pair production in the superconductivity theory).

## Conclusion

We showed that the description of pair production of the electron - positron corresponds to the reformation of the Maxwell linear theory into some nonlinear theory of the electromagnetic field, which is the approximation of the Dirac theory. Such a reformation corresponds to a spontaneous breakdown of the field symmetry in the quantum field theory. We can say that in this case, a spontaneous breakdown of the photon symmetry happens, which corresponds to the creation of two massive particles – the electron and the positron.

The above results show that the Dirac theory can be written in the electromagnetic form as consistently as in the usual (bi)spinor form. But the electromagnetic interpretation of the Dirac theory faces two difficulties:

1) The identification of the Dirac wave function with the electromagnetic field wave function in the case of the linear theory was excluded since the early stage of the development of the quantum mechanics.

It seems to be real that in case of a nonlinear electromagnetic theory the interpretation can be correct.

2) The second difficulty is the following: in the electromagnetic form of the Dirac equation the free term can give three parameters of the electron: the "bare" mass, the "bare" current (charge) and the "bare" radius of the circle electrical current. At the same time the Dirac equation contains only the "bare" mass of the electron; the charge is entered as an additional parameter and the electron radius is equal to zero. As it is known this is the reason why in the QED there are the infinities and the renormalization procedure is needed to remove them.

It can seem, that the conclusions of the nonlinear theory contradict to the results, received in frameworks of the QED. It is not so. QED is the approximation of the nonlinear theory and, hence, it is impossible to compare these theories directly. It is necessary to compare QED with the approximation of the nonlinear theory.

In frameworks of the QED the value $m$ is a free parameter. Obviously, the same parameters within the framework of the nonlinear theory are the current (charge) and radius of a curvature $r_p$, which are connected one to one with $m$. On the basis of the calculations, made within the place of a method of the secondary quantization, the parameter $m$ is equal to the infinity. As the electron charge (parameter $e$) according to the nonlinear theory is proportional to $m$, and the parameter $r_p$ is inversely proportional to $m$, in the frameworks of QED the size of the "charge" will be equal to infinity and the "radius", i.e. the "size" of electron, will be equal to zero. As we know, these both results correspond fully to the results of the QED.

As it is known, to get rid of the infinity in QED the mass and charge renormalization procedure is used. This procedure replaces the parameters $m$ and $e$ with the experimental values of the electron mass and charge: $m_e$ and $e_e$. It is not difficult to show, that in this case we should replace the parameter $r_p$ ("Compton radius") with size "classical radius of electron". (It also explains, why the classical radius of the electron is present in all the first not disappearing members of the formulas of the interaction cross-section of the elementary particles with the electron).

Thus, the linear approximation of the nonlinear theory does not contradict to QED.

It can be supposed that the procedure, which is concurrent to the renormalization procedure, is connected with the introduction of the polarization of the physical vacuum, which will give the solution of the infinity problem as it has place in the known nonlinear electromagnetic theories [5,7].

The representation of the QED in the form of a nonlinear theory of the electromagnetic waves, allows the explanation of many features of the modern quantum theory (see [28] and the following part of the investigation).